\documentclass[aps,pra,twocolumn,superscriptaddress,showpacs,showkeys]{revtex4}

\usepackage{graphicx}
\usepackage{textcomp}
\usepackage{color}
\usepackage{dcolumn}
\usepackage{multirow}
\usepackage{ulem}
\usepackage{amsmath}
\usepackage{amssymb}

\begin{document}

\title{Generating and verifying entangled itinerant microwave fields with efficient and independent measurements}

\author{H.\ S.\ Ku}\email{hsiang-sheng.ku@colorado.edu}
\affiliation{JILA, National Institute of Standards and Technology and The University of Colorado, Boulder, Colorado 80309, USA}
\affiliation{Department of Physics, University of Colorado, Boulder, Colorado 80309, USA}

\author{W.\ F.\ Kindel}
\affiliation{JILA, National Institute of Standards and Technology and The University of Colorado, Boulder, Colorado 80309, USA}
\affiliation{Department of Physics, University of Colorado, Boulder, Colorado 80309, USA}

\author{F. Mallet}
\affiliation{Laboratoire Pierre Aigrain, Ecole Normale Sup\'{e}rieure-PSL Research University, CNRS, Universit\'{e} Pierre et Marie Curie-Sorbonne Universit\'{e}s, Universit\'{e} Paris Diderot-Sorbonne Paris Cit\'{e}, 24 rue Lhomond, 75231 Paris Cedex 05, France}

\author{S. Glancy}
\affiliation{National Institute of Standards and Technology, Boulder, Colorado 80305, USA}

\author{K. D. Irwin}
\affiliation{Department of Physics, Stanford University, Stanford, CA 94305, USA}

\author{G. C. Hilton}
\affiliation{National Institute of Standards and Technology, Boulder, Colorado 80305, USA}

\author{L. R. Vale}
\affiliation{National Institute of Standards and Technology, Boulder, Colorado 80305, USA}

\author{K. W. Lehnert}
\affiliation{JILA, National Institute of Standards and Technology and The University of Colorado, Boulder, Colorado 80309, USA}
\affiliation{Department of Physics, University of Colorado, Boulder, Colorado 80309, USA}

\date{\today}

\begin{abstract}
By combining a squeezed propagating microwave field and an unsqueezed vacuum field on a hybrid (microwave beam-splitter), we generate entanglement between the two output modes. We verify that we have generated entangled states by making independent and efficient single-quadrature measurements of the two output modes. We observe the entanglement witness $E_\mathrm{W}=-0.263^{+0.001}_{-0.036}$ and the negativity $N=0.0824^{+0.01}_{-0.0004}$ with measurement efficiencies at least $26\pm{0.1}\%$ and $41\pm{0.2}\%$ for channel~1 and 2 respectively. These measurements show that the output two-mode state violates the separability criterion and therefore demonstrate entanglement. This shared entanglement between propagating microwaves provides an important resource for building quantum networks with superconducting microwave systems.
\end{abstract}


\maketitle

When two parties share entanglement many powerful quantum communication protocols are available to them. For example, they may communicate with security guaranteed by physical laws, they may encode data more densely than classical bounds, and one party can transfer a quantum state to another by transmitting only classical information, a protocol known as teleportation \cite{Braunstein2005}. Furthermore, teleportation can be extended to realize error correction schemes \cite{aoki2009}. Shared entanglement has been a powerful and popular tool for long distance quantum communications. A second application for shared entanglement occurs in a general quantum information processor that is structured as a distributed machine comprising many well-isolated copies of a high-fidelity quantum register \cite{Dur2003,Jiang2007}. To perform quantum computation, these registers must then share entanglement.

For microwave superconducting qubit circuits, the quantum registers that have the longest coherence time are built from centimeter-sized microwave cavities containing a few qubits \cite{Paik2011}. Propagating microwave modes are the media being developed
to establish and exploit entanglement among such registers. Consequently, entanglement between physically distinct itinerant microwave modes is an important resource and has been created and verified in recent experiments \cite{Flurin2012,Menzel2012,Eichler2012_2,Lang2013}. Although it is possible to verify the presence of entanglement with low efficiency measurements \cite{Menzel2012,Silva2010,Eichler2012_1}, to perform a protocol that exploits shared entanglement between the sender and the receiver, such as teleportation or error correction, having higher detection efficiency improves the fidelity of the process \cite{Braunstein2001,Furusawa1998}. Furthermore, the high-efficiency measurements of the two parties should have independent measurement bases to fully characterize a two-mode state.

For propagating microwave modes the quantities that can be measured with the highest efficiency are quadrature amplitudes $X$ and $Y$, i.e. the cosine and sine components of the field relative to some phase reference. The two quadrature amplitudes are canonically conjugate observables; thus, one quadrature can in principle be measured without added noise, but not both. By adjusting the reference phase, one can measure a linear combination of $X$ and $Y$. We treat $X$ and $Y$ as random variables corresponding to potential outcomes of quadrature measurements.  By making repeated measurements on many copies of the same two-mode state and adjusting their phase references over all possible values, one can fully characterize the two-mode state.

\begin{figure}
 \includegraphics[]{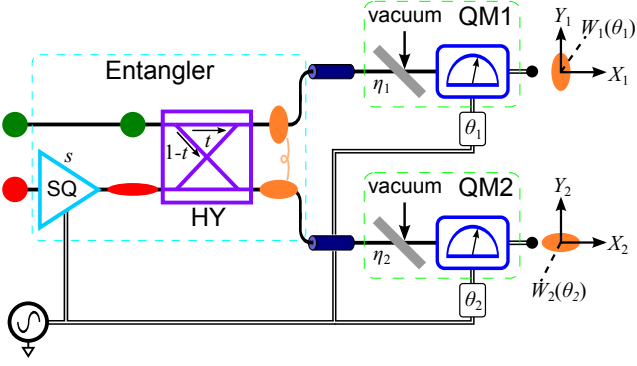}
 \caption{\label{fig_model}~(color online) The single squeezer model of the experiment. The squeezer (SQ) prepares a squeezed state with squeezing parameter $s$, where the variance of the squeezed quadrature is $1/(2s)$. The squeezed state (red ellipse) and the unsqueezed input (green circle) are combined on a quadrature hybrid (HY) to generate entangled modes. The hybrid has a power transmission coefficient $t$ and a power coupling coefficient $1-t$. The two output modes (orange ellipses) of the hybrid propagate onto two physically separate transmission lines and are fed to the two-channel measurement apparatus to measure quadrature amplitudes $W_1(\theta_1)$ and $W_2(\theta_2)$. The measurement apparatus consists of two single-quadrature measurement chains (QM1 and QM2), where each QM employs a VER as the first amplifier. All sources of loss (including loss inside the SQ) and measurement inefficiencies are modeled by introducing two fictitious beam splitters with power transmission coefficients $\eta_1$ and $\eta_2$, respectively. The two squeezed states arrive at the two VERs with fixed but uncontrolled phase shifts. We mathematically adjust the reference phases to align the squeezed states with $X_1$ and $X_2$ as illustrated.}
\end{figure}

In this paper, we report the generation of entanglement between two spatially separate itinerant microwave modes by combining a quadrature squeezed state and a vacuum state on a microwave hybrid (Fig.~\ref{fig_model}). A similar method to create entanglement is commonly used in optical experiments, such as \cite{Eberle2011}. To verify entanglement, we make efficient, single-quadrature measurements simultaneously on the two separate modes with independent control of phases $\theta_1$ and $\theta_2$. Specifically, we measure quadrature amplitudes of the two modes $W_1(\theta_1)=X_1\cos(\theta_1)+Y_1\sin(\theta_1)$ and $W_2(\theta_2)=X_2\cos(\theta_2)+Y_2\sin(\theta_2)$ over all pairs $(\theta_1, \theta_2)$. Because the two-mode state in this experiment appears to be Gaussian, we characterize it using its quadrature covariance matrix defined as $\Sigma_{ij}=(1/2)\langle Z_iZ_j+Z_jZ_i\rangle-\langle Z_i\rangle\langle Z_j\rangle$, where $Z_i\in\{X_1,Y_1,X_2,Y_2\}$ \cite{Simon1994}.

The covariance matrix reveals the correlations between the two modes and can be used to demonstrate entanglement. Specifically, one can prove entanglement by observing violation of the inequality \cite{Duan2000,Simon2000,Bowen2004}
\begin{eqnarray}
R(\theta_1,\theta_2,a) = \mathrm{Var}\left[\left|a\right|W_1(\theta_1)+\frac{1}{a}W_2(\theta_2)\right] + \nonumber \\ \mathrm{Var}\left[\left|a\right|W_1(\theta_1+\frac{\pi}{2})-\frac{1}{a}W_2(\theta_2+\frac{\pi}{2})\right] \nonumber \\ \geq \left(a^2+\frac{1}{a^2}\right),\label{sepaCrit}
\end{eqnarray}
for any nonzero real number $a$, which accounts for any unbalance between the two channels. (We use the convention that for vacuum states $\mathrm{Var}(W(\theta))=1/2$ for all $\theta$.) One may optimize the observed violation over phase rotations (which cannot change entanglement) and $a$ by computing the entanglement witness
\begin{equation} 
E_{\mathrm{W}}~=~\min_{\theta_1,\theta_2,a} \left[R(\theta_1,\theta_2,a)-\left(a^2+\frac{1}{a^2}\right)\right],\label{Ew}
\end{equation}
so $E_\mathrm{W} < 0$ is evidence of entanglement. For the choice of $a=1$, we also report $\Delta_\mathrm{EPR}=(1/2)R(\theta_1,\theta_2,1)$ which gives evidence of entanglement when $\Delta_\mathrm{EPR} < 1$. Although the entanglement witness can detect the presence of entanglement, it does not measure the amount of entanglement. Thus, we quantify the entanglement between the two measured microwave modes with the negativity $N$, which can also be calculated from the covariance matrix \cite{Vidal2002,Adesso2004}. For bipartite systems, the negativity is a lower bound on the number of entangled Hilbert space dimensions (the Schmidt number) \cite{Eltschka2013}.

As is evident from Eq.~(\ref{sepaCrit}), two vacuum modes are not entangled. We must use some device to actively transform the vacuum. The Josephson Parametric Amplifier (JPA) is the critical piece of technology that allows us to generate squeezed states and to perform efficient quadrature measurements at microwave frequencies \cite{Mallet2011}. It is a microwave phase-sensitive amplifier that is built from an electrically nonlinear microwave resonant circuit and that derives its gain from a pump tone exciting the circuit. The JPA amplifies a specific quadrature of the input state noiselessly while squeezing the conjugate quadrature \cite{Yurke1988,Yurke1989}, where the amplified quadrature is selected by the relative phase between the pump tone and the input state. In this experiment, one JPA denoted SQ is used to transform the vacuum to a squeezed state. Two more JPAs, denoted VER1 and VER2, act as single-quadrature preamplifiers for two microwave measurement chains. By adjusting the pump phases of VER1 and VER2 separately, we independently control the bases of our two single-quadrature measurements.

In this experiment, we integrate SQ with a hybrid on a single chip forming the entangler circuit \cite{Ku2011} (see supplementary for the layout and an image of the entangler) to minimize loss in the entanglement generation process. Feeding the two inputs of the entangler are two vacuum states emitted from two $50~\Omega$ terminations thermally anchored to a cryostat. SQ is pumped at $f_\mathrm{s}=6.327~\mathrm{GHz}$ and generates a squeezed state with approximate direct power gain $G_\mathrm{s}=3.1~\mathrm{dB}$ and bandwidth $B_\mathrm{s}=8.5~\mathrm{MHz}$. The squeezed state is displaced in phase space by the pump amplitude. This displaced squeezed state and the vacuum state interfere in the hybrid, creating entanglement in the two output modes of the hybrid. In order to ensure that the following VERs are not saturated, we null the SQ pump tone at the input of the VERs with a weakly coupled coherent field. The SQ's gain is chosen as a compromise between larger observed squeezing and simpler operation of the experiment. SQ gain greater than 5~dB will not substantially reduce the measured variance because of the measurement inefficiency, but the associated large pump amplitude will be more difficult to null at the VERs' inputs.

The two-mode state is then measured by our two-channel measurement apparatus. The two output modes of the entangler propagate in two separate coaxial cables which feed the input of VER1 or VER2, located about 10~cm apart, forming the first stages of the amplification of the two quadrature measurements. The VERs are operated with approximate direct power gain $G_\mathrm{v}=22~\mathrm{dB}$ and bandwidth $B_\mathrm{v}=2.3~\mathrm{MHz}$, and their outputs are further amplified by conventional microwave amplifiers. The VERs' gains are chosen to be large enough to overwhelm the added noise of the following HEMT amplifiers (20~dB) but not larger as that would reduce the measurement bandwidth. Each of the three JPAs' gains is estimated by measuring its response to a small input tone, but with all of other JPAs turned off. Because the JPA gains may change by about $2\%$ when all three JPAs are turned on, these are only estimates of the gains during the entanglement generation and verification.

Finally, the amplified microwave signals of the two channels are mixed down for digitizing with copies of the VERs' pump tones serving as the mixers' local oscillators. The mixers' intermediate frequency outputs are filtered with a 1.9~MHz low pass filter and sampled at 10~MHz, yielding new measurements of $W_1(\theta_1)$ and $W_2(\theta_2)$ every 100~ns. The phases $\theta_1$ and $\theta_2$ can be independently adjusted relative to each other and relative to the squeezed quadrature of SQ. In practice, to adjust $\theta_1$ and $\theta_2$, we set the pump frequencies of VER1 and VER2 to be 1~kHz and 50~kHz above the SQ pump frequency respectively. In 1~ms, we acquire 10,000 samples covering the full range of both $\theta_1$ and $\theta_2$. We acquire data for 1~s, yielding 1,000 independent realizations of ($W_1$, $W_2$) for each pair of ($\theta_1$, $\theta_2$).
\begin{figure}
 \includegraphics[]{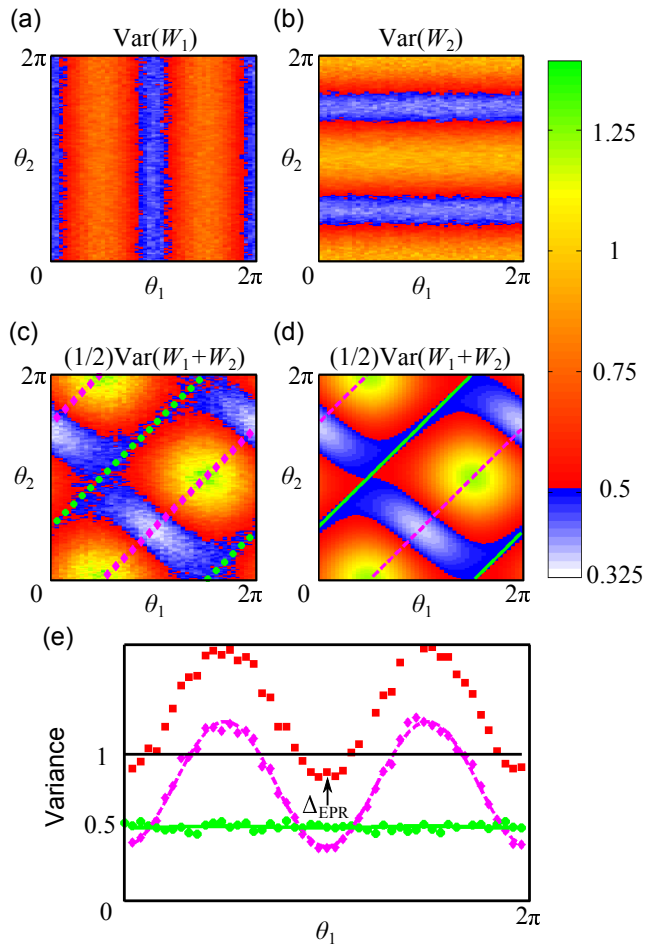}
 \caption{\label{fig_vars}~(color online) Separate and joint variances. Shown are intensity plots of the measured variances of (a) $W_1(\theta_1,\theta_2)$, of (b) $W_2(\theta_1,\theta_2)$, and of (c) $\frac{1}{2}\left[W_1(\theta_1,\theta_2)+W_2(\theta_1,\theta_2)\right]$ calibrated in units of the vacuum versus the two quadrature phases $\theta_1$ and $\theta_2$. (d) An expectation of (c) predicted by the single squeezer model represented by Fig.~\ref{fig_model}. (e) The variances along the corresponding annotated lines in (c) and (d) are plotted versus quadrature phase $\theta_1$ of channel 1. The red squares are the sum of the green-circle line and the magenta-diamond line, where the green-circle line is shifted by $\frac{\pi}{2}$ in $\theta_1$. The arrow indicates the observed value of $\Delta_\mathrm{EPR}<1$.}
\end{figure}

To calibrate our measured variances in units of the vacuum, we inject states of known noise into the entangler, but we bypass SQ by turning off its pump and operating it as a noiseless $G_\mathrm{s}=1$ amplifier. States of known noise are created by varying the temperature of the cryostat and therefore the $50~\Omega$ terminations that feed the entangler. We then measure the variances at the outputs of the measurement chains while adjusting the cryostat temperature, thus calibrating the measurement chains. From this procedure, we also determine that the $50~\Omega$ terminations equilibrate with the cryostat for any temperature above $25~\mathrm{mK}$ (see Appendix~\ref{Appen_A}). Because the entanglement generation is performed with the cryostat below 25~mK, the input variance is indistinguishable from vacuum in our experiments.

In order to interpret the measurements, we construct an analytic single squeezer model of the experiment represented by Fig.~\ref{fig_model}. All sources of loss and measured inefficiencies are absorbed into the parameters $\eta_1$ and $\eta_2$. We extract parameters by fitting the model to the measured variances yielding $s=5.41\pm{0.03}$, $\eta_1t=0.130\pm{0.001}$, and $\eta_2(1-t)=0.202\pm{0.001}$. Moreover, we are able to extract from the model small changes in the VERs' gains when bypassing SQ (see Appendix~\ref{Appen_A}). We conclude that VER1's and VER2's power gains change by linear factors of approximately $-1.70\pm{0.07}\%$ and $2.04\pm{0.08}\%$ respectively when the SQ pump is turned on.

To test the independence between the two measurement channels, we first examined the quadrature variances measured separately at the outputs of the two measurement chains. From the measured quadrature variance of each mode, $\mathrm{Var}(W_1)$ and $\mathrm{Var}(W_2)$ [Fig.~\ref{fig_vars}(a) and (b)], we observe an approximate minimum variance $15\%$ below vacuum fluctuation, i.e.\ $15\%$ squeezing below vacuum. Furthermore, $\mathrm{Var}(W_1)$, only depends on the measurement phase $\theta_1$ and is independent of the measurement phase $\theta_2$; likewise $\mathrm{Var}(W_2)$ depends only on $\theta_2$. These plots demonstrate that the two VERs are unaffected by the phases of the other's pump indicating that the two channels are well decoupled. Finally, the $\pi$ rather than $2\pi$ periodicity of $\mathrm{Var}(W_1)$ [$\mathrm{Var}(W_2)$] as a function of $\theta_1$ ($\theta_2$) shows that the SQ pump is successfully nulled at the input of the VERs.

By making joint measurements of the two output modes, we detect the correlations between them and reveal that they are entangled. In Fig.~\ref{fig_vars}(c) we plot the measured joint variance $(1/2)\mathrm{Var}(W_1+W_2)$, and in Fig.~\ref{fig_vars}(d) we show the expected joint variance predicted by the single squeezer model in Fig.~\ref{fig_model}. Because the squeezing we observed from separate measurements is diluted with vacuum (Fig.~\ref{fig_model}), we anticipate that the joint measurements will show more squeezing than the separate measurements. Indeed, $(1/2)\mathrm{Var}(W_1+W_2)$ has an approximate minimum variance $25\%$ below vacuum fluctuation. The hybrid generates a two-mode entangled state, distributing the squeezing present in the input squeezed state into the two output modes. By mathematically inverting the hybrid's action, the two input states can be reconstructed from the joint measurements. For example, only the squeezed input contributes to the variance measured along the magenta-diamond line in Fig.~\ref{fig_vars}(c); likewise, only the vacuum input contributes to the variance measured along the green-circle line. One method of estimating $\Delta_\mathrm{EPR}$ is to sum the variances measured along these two lines [Fig.~\ref{fig_vars}(e)]. To see this, note that the second term in the expression for $\Delta_{\mathrm{EPR}}$ can be written as (1/2)$\mathrm{Var}\left[W_1(\theta_1^\prime)+W_2(\theta_2^\prime)\right]$, where $\theta_1^\prime = \theta_1 - \pi/2$ and $\theta_2^\prime = \theta_2 + \pi/2$. Thus, if the first term in $\Delta_{\mathrm{EPR}}$ is evaluated at ($\theta_1$,$\theta_2$) on the magenta-diamond line, then the second term must be evaluated at a corresponding point ($\theta_1^\prime$,$\theta_2^\prime$) along the green-circle line. By direct inspection of the joint variance, this $\Delta_\mathrm{EPR}<1$ already suggests that the two modes are entangled.
\begin{figure}
 \includegraphics[]{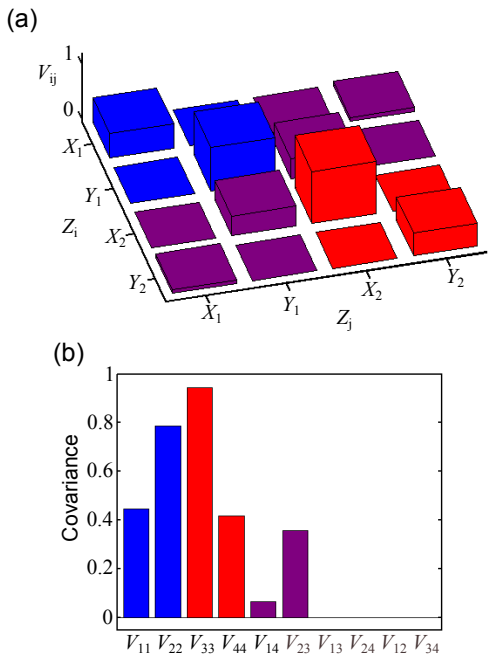}
 \caption{\label{fig_cov}~(color online) Covariance matrix of the two-mode state. (a) The covariance matrix calculated from the single squeezer model parameter extracted by joint fitting of measured variances. (b) The 10 independent elements of the covariance matrix are shown for the same data as (a).}
\end{figure}

Instead of extracting $E_\mathrm{W}$ from a particular point in Fig.~\ref{fig_vars}(e), we can reduce uncertainty by calculating $E_\mathrm{W}$ by using the quadrature measurements made over all $(\theta_1, \theta_2)$. By fitting all the measured variances to the single squeezer model represented in Fig.~\ref{fig_model}, we calculate the covariance matrix from extracted model parameters (see Appendix~\ref{Appen_B}). Consistent with the measured variances (Fig.~\ref{fig_vars}), the elements of the covariance matrix show both modest squeezing within each mode, and stronger intermode correlations (Fig.~\ref{fig_cov}). From the covariance matrix, we directly calculate the entanglement witness $E_\mathrm{W}=-0.263\pm{0.001}$ with $a=1.11$. To quantify the degree of entanglement, we also calculate the negativity $N=0.0824\pm{0.0004}$. The uncertainties of $E_\mathrm{W}$ and $N$ are estimated from a parametric bootstrap method. In the bootstrap procedure, 20 simulated data sets are generated from the measured covariance matrix. We estimated statistical uncertainties as equal to the standard deviation of the 20 estimates $E_\mathrm{W}$ and $N$ found by analyzing the simulated data sets in the same manner used for analyzing real data. Furthermore, to demonstrate the repeatability and stability of the entanglement generation, we perform 100 trials of the experiment. (The quoted $E_\mathrm{W}$ and $N$ are one typical result from 100 trials of the experiment.) The distribution of $E_\mathrm{W}$ and $N$ calculated from 100 trials appear to be Gaussian with a mean $\pm$ standard~deviation of $E_\mathrm{W}=-0.264\pm{0.002}$ and $N=0.0820\pm{0.0007}$.

Because we acquire a large data set in a short time, the statistical uncertainties of $E_\mathrm{W}$ and $N$ are low compared to the systematic errors. We investigate systematic errors by analyzing experimental data with a general Gaussian model (see Appendix~\ref{Appen_C}). Whereas the single squeezer model restricts possible estimated covariance matrices to those generated by a single mode squeezer, beam splitter, and loss, the Gaussian model allows any two mode state described by a Gaussian Wigner function. Using this Gaussian model, we found that $E_\mathrm{W}=-0.297\pm0.002$ and $N=0.0921\pm0.0004$, where the uncertainties are based on a similar parametric bootstrap method to that described above. The discrepancy between the parameters estimated by the single squeezer model and the Gaussian model reveals systematic error that is significantly larger than any statistical uncertainty and uncertainty in the variances of the calibration state. Considering this systematic error, we conclude that the state in our experiment has $E_\mathrm{W}=-0.263^{+0.001}_{-0.036}$ and $N=0.0824^{+0.01}_{-0.0004}$, where the uncertainties include the estimates from both models.

The observed negativity is small compared to $N=0.61$ and $N=0.55$ from \cite{Flurin2012} and \cite{Menzel2012}, respectively, but those negativities are inferences of the negativity in the absence of measurement inefficiencies. However, reference \cite{Flurin2012} effectively combines two squeezed states on a hybrid, therefore, it is possible to generate states with larger negativity using that method than the method demonstrated here and in \cite{Menzel2012}. We state the negativity without correcting for measurement inefficiencies. In comparison to the noisy amplification used in \cite{Menzel2012} and the approximately $14\%$ efficiency achieved in \cite{Flurin2012}, our apparatus achieves measurement efficiencies at least $\eta_1=26\pm{0.1}\%$ and $\eta_2=41\pm{0.2}\%$ (To quote these efficiencies, we assume $t=0.51$ based on the calibrated measurements of the hybrid \cite{Ku2011}). Both this work and \cite{Flurin2012} benefit from quantum-efficient preamplifiers. The efficiencies presented here are higher than in \cite{Flurin2012} simply because we operate our preamplifiers (VERs) with higher gains further reducing the apparent noise added by the following HEMT amplifiers. The efficiencies include any noise in the squeezed state's generation and any loss along the entire path from state generation to measurement, in other words, the quoted values are lower bounds of the detection efficiencies. Because we know only lower bounds on the detection efficiencies we are unable to estimate the negativity (or other properties) of the generated state in the absence of measurement inefficiency.

In conclusion, we demonstrate a two-channel, single-quadrature quantum measurement apparatus in the microwave regime, where each channel of the apparatus uses a JPA as its first stage amplifier. A two-mode entangled state, which is generated by combining a squeezed state and vacuum on a microwave hybrid, is measured with improved efficiency and independent choices of each mode's measured quadratures. Entanglement is demonstrated by showing that the two-mode state violates the separability criterion. Our integration of JPAs for both the preparation and measurement of an entangled state is a substantial addition to the toolbox for manipulating continuous variable quantum states of microwave modes. The measurement scheme is promising for demonstrating protocols exploiting entanglement.

\appendix

\section{Vacuum calibration}\label{Appen_A}
\begin{figure*}
 \centering
 \includegraphics[width=6.5in]{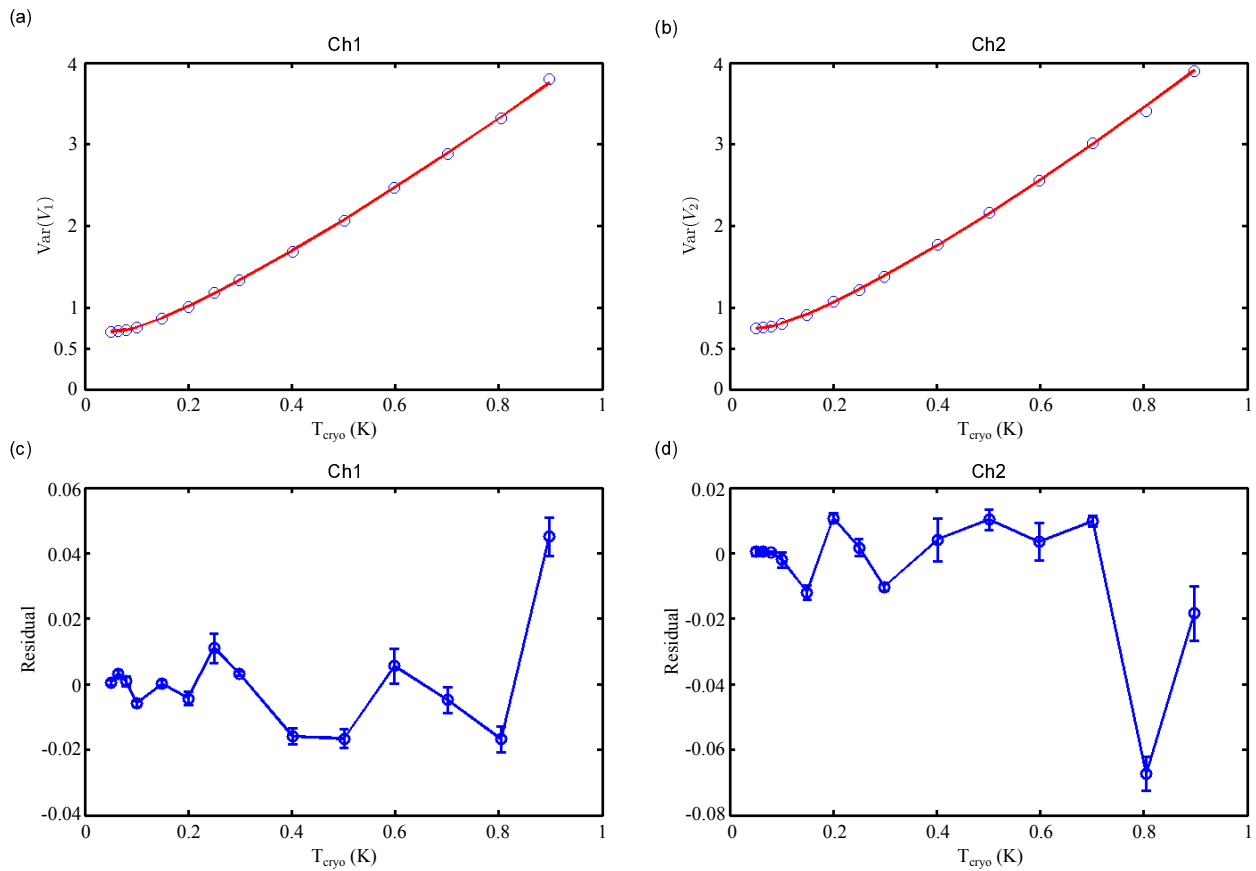}
 \caption{\label{fig_ThermalSweep}~Thermal sweep experiment. The measured single-quadrature variances of the thermal sweep experiment (blue circles) and fit to Eq.~(\ref{themalsweep}) (red solid line) for channel 1 (a) and channel 2 (b) are plotted. The residuals of the two fits are plotted in (c) and (d) where the error bars show the standard deviation of five independent measurements made at each temperature point.}
\end{figure*}

When we demonstrate the two-mode entanglement experiment, we feed the entangler with two input states emitted from the two $50~\Omega$ terminations thermally anchored to the cryostat. In additional to these two input states, more thermal (nearly vacuum) modes dilute the entangled state through the losses of the commercial microwave components, such as directional couplers and circulators. The temperatures of the input states and the loss modes are assumed to be equal to $T_\mathrm{in}$. To measure $T_\mathrm{in}$, we inject a series of known thermal states into the two measurement chains without pumping SQ and calculate the variances $\mathrm{Var}(V_1)$ and $\mathrm{Var}(V_2)$ of the two output measurements for each thermal input. We then fit the model
\begin{eqnarray}
\mathrm{Var}(V) = G\left[\frac{1}{2}\coth\left(\frac{hf_s}{2k_\mathrm{B}T_{\rm_{in}}}\right)+A(T_\mathrm{F})\right],\label{themalsweep}
\end{eqnarray}
to the measured variances, where $G$ and $A(T_\mathrm{F})$ are the power gain and the added noise of the measurement chain. In this model, $T_\mathrm{in}=\sqrt{T_\mathrm{F}^2+T_\mathrm{e}^2}$ represents the input states temperature. $T_\mathrm{F}$ is the cryostat temperature, and we add one parameter $T_\mathrm{e}$ to allow for the possibility that the terminations equilibrate at a higher temperature than $T_\mathrm{F}$. We also include a temperature dependent added noise $A(T_\mathrm{F})=A_0+A_2T_\mathrm{F}^2$ in the model. The origin of the temperature dependent added noise in our JPAs is still under investigation, but it seems to be caused by the presence of resistive filters in the on-chip bias lines. From the fit (Fig.~\ref{fig_ThermalSweep}), we extract $0<T_\mathrm{e}<16.1$~mK. Because the entanglement generation is performed with the cryostat below 25~mK, we have $T_\mathrm{in}<29.7$~mK, which means the input variance is indistinguishable from vacuum in our experiments.

\begin{figure*}
 \centering
 \includegraphics[width=6.5in]{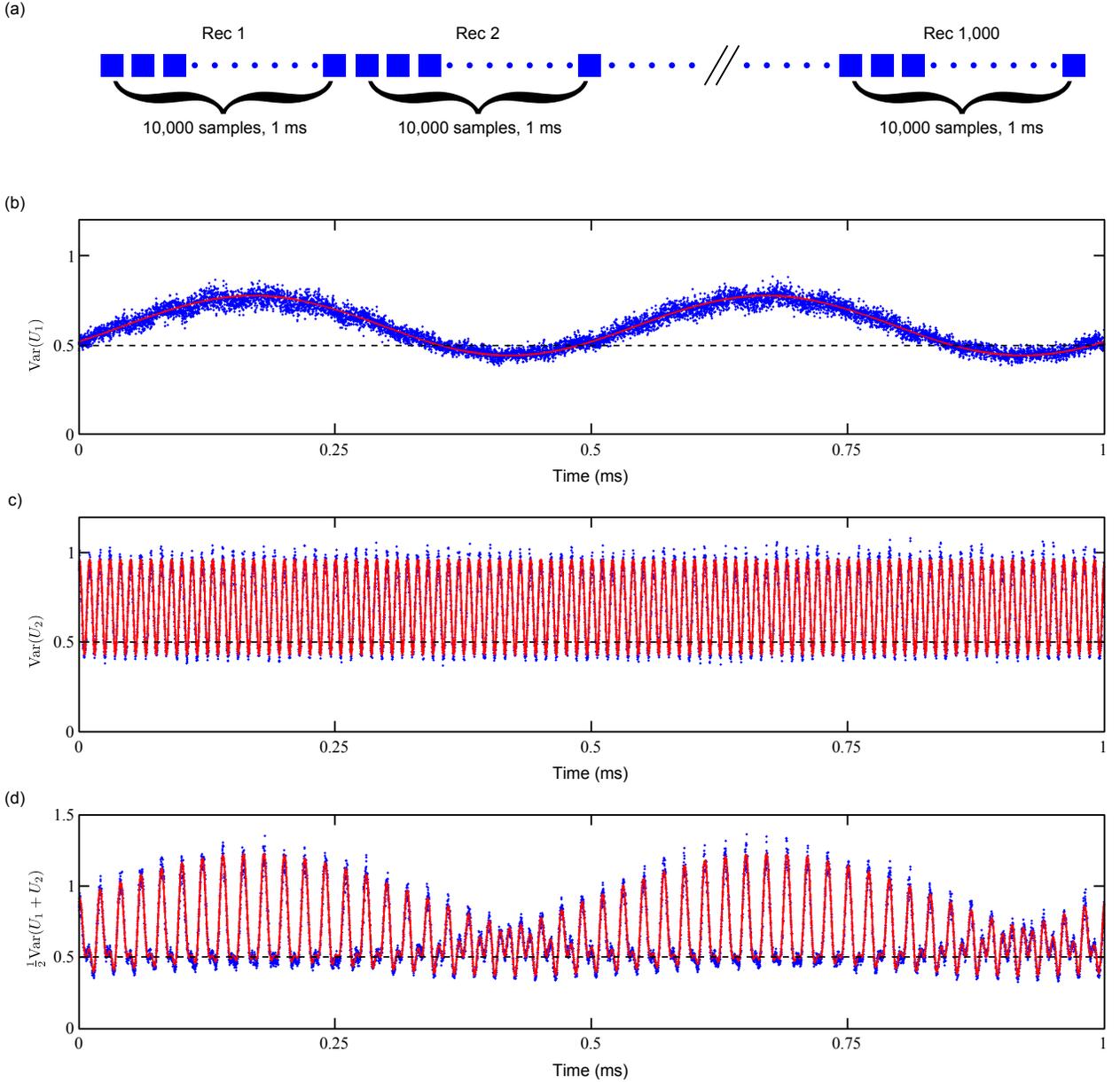}
 \caption{\label{fig_varsfit}~ Joint fit of the model equations to the measured variances. (a) The cartoon diagram represents the timing of the quadrature measurements. Each measurement is a one second long time trace with each channel sampled at 10~MSamples/s. Because the pumps of VER1 and VER2 are detuned from the SQ pump by 1~kHz and 50~kHz, respectively, the two VERs amplify the measured states at 10,000 different quadrature phase combinations every 1~ms (one record). In a 1~s long time trace, we thus have 1000 records, or realizations at each quadrature phase pair. The variance at each quadrature phase pair is then calculated from the corresponding data points in each record. The measured variances of (b) $U_1(\theta_1)$, of (c) $U_2(\theta_2)$, and of (d) $1/2\left[U_1(\theta_1)+U_2(\theta_2)\right]$ are plotted versus time in each measurement record (blue dot) along with the joint fit of Eq.~(\ref{var1})--(\ref{var1and2}) to all of the measured variances (red solid line). The model and data show good agreement with each other and we observe moderate squeezing below vacuum fluctuation (0.5) in the variance of each mode [(b) and (c)], but more squeezing in the joint measurement (d).}
\end{figure*}

The experiment's data set contains amplified quadrature measurements $V_{i,\mathrm{off}}$  (measured by bypassing SQ) of the vacuum state and quadrature measurements $V_{i,\mathrm{on}}$ of the entangled state for modes $i=1,2$.  Those variances, measured by digitizing the amplified voltages at room temperature, must be calibrated in units of quantum vacuum at the input of the experiment. To calibrate the variances of the measurements in units of vacuum, we first normalize the quadrature measurements of the two entangled modes by the variances of the vacuum states to get $U_1(\theta_1)$ and $U_2(\theta_2)$
\begin{equation*}
U_1(\theta_1)=\frac{V_{1,\mathrm{on}}(\theta_1)}{\sqrt{\mathrm{Var}(V_{1,\mathrm{off}})}},\quad U_2(\theta_2)=\frac{V_{1,\mathrm{on}}(\theta_2)}{\sqrt{\mathrm{Var}(V_{2,\mathrm{off}})}}.
\end{equation*}
Furthermore, we derive the equations for variances of $U_1(\theta_1)$, of $U_2(\theta_2)$, and of $U_1(\theta_1) \pm U_2(\theta_2)$ from the single-squeezer model
\begin{widetext}
\begin{eqnarray}
\mathrm{Var}\left[U_1(\theta_1)\right]&=&g_1\left[1 + \frac{1}{2}\frac{(s-1)^2}{s}\alpha+\frac{1}{2}\frac{s^2-1}{s}\alpha\cos(2\theta_1+2\phi_1)\right],\label{var1}\\
\mathrm{Var}\left[U_2(\theta_2)\right]&=&g_2\left[1 + \frac{1}{2}\frac{(s-1)^2}{s}\beta+\frac{1}{2}\frac{s^2-1}{s}\beta\cos(2\theta_2+2\phi_2)\right],\label{var2}\\
\mathrm{Var}\left[U_1(\theta_1) \pm U_2(\theta_2)\right]&=&\mathrm{Var}(U_1)+\mathrm{Var}(U_2) \nonumber \\
\pm\sqrt{g_1g_2}&\left[\right.&\frac{s^2-1}{s}\sqrt{\alpha\beta}\cos(\theta_1+\theta_2+\phi_1+\phi_2)+\frac{(s-1)^2}{s}\sqrt{\alpha\beta}\cos(\theta_2-\theta_1+\phi_2-\phi_1)\left.\right].\label{var1and2}
\end{eqnarray}
\end{widetext}
The parameters $\alpha$ and $\beta$ combine the hybrid power coupling coefficient $t$ and the two measurement efficiencies $\eta_1$ and $\eta_2$
\begin{equation*}
\alpha=t\eta_1 \quad\mathrm{and}\quad \beta=(1-t)\eta_2.
\end{equation*}
We do not assume that the VERs are perfectly linear, but rather introduce two parameters $g_1$ and $g_2$ to model the changes in VERs' gains when bypassing SQ:
\begin{equation*}
g_1 = \frac{g_{\rm{1,on}}}{g_{\rm{1,off}}}, \quad g_2 = \frac{g_{\rm{2,on}}}{g_{\rm{2,off}}},
\end{equation*}
where $g_{\rm{1,on}}$ and $g_{\rm{1,off}}$ are quadrature power gains of mode~1 for SQ is operated or bypassed and the same for mode~2. Finally, from the joint fit of the model equations (\ref{var1})--(\ref{var1and2}) to the measured $\mathrm{Var}(U_1)$, $\mathrm{Var}(U_2)$, and $\mathrm{Var}(U_1 \pm U_2)$ [Fig.~\ref{fig_varsfit}(b)--(d)], we extract $s=5.41\pm{0.03}$, $\alpha=0.1304\pm{0.0007}$, $\beta=0.202\pm{0.001}$, $\phi_1=-1.070\pm.0.002$, $\phi_2=-0.176\pm0.001$, $g_1=-1.70\pm{0.07}~\%$, and $g_2=2.04\pm{0.08}~\%$. We then are able to use $g_1$, $g_2$, and $T_\mathrm{in}$ to calibrate the quadrature measurements in units of vacuum:
\begin{equation*}
W_1(\theta_1)=\frac{U_1(\theta_1)}{\sqrt{g_1}}\frac{\sigma}{0.5}, \quad
W_2(\theta_2)=\frac{U_2(\theta_2)}{\sqrt{g_2}}\frac{\sigma}{0.5},
\end{equation*}
where $\sigma=(1/2)\coth(hf_s/2k_\mathrm{B}T_{\rm_{in}})$.

\section{Covariance matrix, the entanglement witness, and the negativity}\label{Appen_B}
To compute the covariance matrix predicted by the single-squeezer model, we examine the effects of each of the linear optical transformations (including squeezing) shown in Fig. 1 of the main text. Each of the transformations evolves the quadrature vector $Z=\{X_1, Y_1, X_2, Y_2\}^{\text{T}}$ to $MZ$. The matrix $M$ that describes the transformation is a real $4\times4$ matrix in the symplectic group $\mathrm{Sp}(4,\mathbb{R})$ \cite{Simon1994}. The quadratures of the transformed state will have the covariance matrix $M\Sigma M^{\text{T}}$, where $\Sigma$ was the covariance matrix of the original state.  The squeezing of mode 1 by amount $s$ is described by
\begin{equation*}
S(s) =
\begin{bmatrix}
\frac{1}{\sqrt{s}} & 0 & 0 & 0 \\
0 & \sqrt{s} & 0 & 0 \\
0 & 0 & 1 & 0 \\
0 & 0 & 0 & 1
\end{bmatrix}.
\end{equation*}
A beam splitter of transmissivity $t$ is described by
\begin{equation*}
B(t) =
\begin{bmatrix}
\sqrt{t} & 0 & -\sqrt{1-t} & 0 \\
0 & \sqrt{t} & 0 & -\sqrt{1-t} \\
\sqrt{1-t} & 0 & \sqrt{t} & 0 \\
0 & \sqrt{1-t} & 0 & \sqrt{t}
\end{bmatrix}.
\end{equation*}
Phase shifting mode 1 by $\phi_1$ and mode 2 by $\phi_2$ is described by
\begin{equation*}
P(\phi_1, \phi_2) =
\begin{bmatrix}
\cos(\phi_1) & \sin(\phi_1) & 0 & 0 \\
-\sin(\phi_1) & \cos(\phi_1) & 0 & 0 \\
0 & 0 & \cos(\phi_2) & \sin(\phi_2) \\
0 & 0 & -\sin(\phi_2) & \cos(\phi_2)
\end{bmatrix}.
\end{equation*}
Beginning with an initial vacuum state, $\Sigma_0=\mathbb{I}/2$, (with $\mathbb{I}$ being the identity matrix), we compute the state just prior to measurement to be $\Sigma_p=P(\phi_1,\phi_2)B(t)S(s)\Sigma_0S(s)^{\text{T}}B(t)^{\text{T}}P(\phi_1,\phi_2)^{\text{T}}$. To account for photon loss we append two ancilla modes, which are coupled to modes 1 and 2 with beam splitters $B(\eta_1)$ and $B(\eta_2)$ respectively. After this coupling, the ancilla modes are discarded. This procedure transforms $\Sigma_p$ into
\begin{equation*}
\Sigma_{\text{ss}} = H\Sigma_p + \bar{H}\frac{\mathbb{I}}{2},
\end{equation*}
where $H=\mathrm{diag}(\eta_1,\eta_1,\eta_2,\eta_2)$ and $\bar{H}=\mathrm{diag}(1-\eta_1,1-\eta_1,1-\eta_2,1-\eta_2)$.

Given a covariance matrix $\Sigma$, we compute the value of the entanglement witness $E_{\text{W}}$ (shown in Eq.~(\ref{Ew}) of the main text) by
\begin{widetext}
\begin{equation*}
E_{\text{W}} = \min_{\theta_1, \theta_2, a}\left[ {a_x}^{\text{T}}P(\theta_1,
\theta_2)\Sigma P(\theta_1, \theta_2)^{\text{T}}a_x +
{a_y}^{\text{T}}P(\theta_1, \theta_2)\Sigma P(\theta_1, \theta_2)^{\text{T}}a_y
- \left(a^2+\frac{1}{a^2}\right)\right],
\end{equation*}
\end{widetext}
where ${a_x}^{\text{T}}=(|a|, 0, 1/a, 0)$ and ${a_y}^{\text{T}}=(0, |a|,
0, -1/a)$.  In fact, $E_{\text{W}}$'s only dependence on $\theta_1$ and
$\theta_2$ appears in the form $\theta_1 - \theta_2$, so it is only
necessary to minimize over one phase.

To compute the entanglement negativity from a covariance matrix, we follow the treatment given in \cite{Adesso2004}. Through the application of linear optical devices and squeezing (the symplectic transformations), any 2 mode covariance matrix $\Sigma$ can be transformed into the covariance matrix of a thermal state, which has the form $\text{diag}(\nu_1, \nu_1, \nu_2, \nu_2)$.  $\nu_1$ and $\nu_2$ are called the symplectic eigenvalues of $\Sigma$. According to the Heisenberg Uncertainty Principle, $\nu_1$ and $\nu_2 \geq 1/2$. If the quantum state $\rho$ with covariance matrix $\Sigma$ is separable, then covariance matrix $\tilde{\Sigma}$ of the partial transpose of $\rho$ will also have symplectic eigenvalues $\tilde{\nu_1}$ and $\tilde{\nu_2} \geq 1/2$.  If $\tilde{\nu_1}$ or $\tilde{\nu_2} < 1/2$, $\rho$ must be entangled.  To compute the symplectic eigenvalues of $\tilde{\Sigma}$, we divide $\Sigma$ into $2\times 2$ blocks:
\begin{equation*}
\Sigma=\begin{pmatrix}
A & \Gamma \\
\Gamma^{\text{T}} & B.
\end{pmatrix}
\end{equation*}
The quantities $|\Sigma|$ and $\Delta(\Sigma)=|A|+|B|+2|\Gamma|$ are invariant under the symplectic transformations.  ($|\cdot|$ denotes the determinant.) From them we calculate the symplectic eigenvalues:
\begin{equation*}
\nu_i = \sqrt{\frac{1}{2}\left(\Delta(\Sigma) \pm \sqrt{[\Delta(\Sigma)]^2-4|\Sigma|}\right)},
\end{equation*}
where we use $i=1$ for the ${}-{}$ case and $i=2$ for the ${}+{}$ case.  The partial transposition of $\rho$ has the effect of reversing the sign of the $Y$ quadrature of the transposed mode, so that $\Delta(\tilde{\Sigma})=|A|+|B|-2|\Gamma|$, but $|\tilde{\Sigma}|=|\Sigma|$. The symplectic eigenvalues of $\tilde{\Sigma}$ are
\begin{equation*}
\tilde{\nu_i} = \sqrt{\frac{1}{2}\left(\Delta(\tilde{\Sigma}) \pm \sqrt{[\Delta(\tilde{\Sigma})]^2-4|\Sigma|}\right)},
\end{equation*}
If $\tilde{\nu_1} < 1/2$, $\rho$ is an entangled state. Finally, the negativity is
\begin{equation*}
N=\max\left(0, \frac{\frac{1}{2}-\tilde{\nu_1}}{2\tilde{\nu_1}}\right).
\end{equation*}
Note that our formula for $N$ is slightly different from that in \cite{Adesso2004}, because that paper uses the convention that the variance of the vacuum state is $1$, whereas we use vacuum variance of $1/2$.

\section{Gaussian state estimation}\label{Appen_C}
As a check for systematic errors we implemented a second method to estimate the quantum state produced in this experiment. We call this method ``Gaussian state estimation'', because its range is all two-mode states that have Gaussian Wigner functions. As inputs it accepts the calibrated quadrature measurements $W_i(\theta_i)$ for $i=1,2$ and returns the Gaussian state's vector of quadrature expected values $\mu = (\mu_{x1},\mu_{y1},\mu_{x2},\mu_{y2})^T = \mathrm{E}[(X_1,Y_1,X_2,Y_2)^T]$ ($\mathrm{E}[x]$ is the expectation value of $x$.) and the covariance matrix $\Sigma$ whose elements are the covariances between of the random variables $(X_1,Y_1,X_2,Y_2)^T$. We will label the elements of $\Sigma$ with subscripts indicating the quadrature variable and mode: $\Sigma_{ai,bj}$, where $a$ and $b\in\{x,y\}$ and $i$ and $j\in\{1,2\}$.  Note that the Gaussian state estimation requires the use of the squeezer model described in Appendix~\ref{Appen_A} to produce correctly calibrated quadrature measurements.  The data set consists of the $n$ instances of the quadruplet containing two phases and two quadrature measurements: $\{(\theta_1^{(k)}, W_1^{(k)}, \theta_2^{(k)}, W_2^{(k)})|k=1,...,n\}$ in which $k$ labels the measurement instance. Although in the experiment $\theta_1$ and $\theta_2$ are scanned continuously and quadratures are measured at regular intervals, in this section we treat the phases as random variables uniformly distributed over $[0,2\pi)$ with probability distribution $P(\theta_i)=1/(2\pi)$.

Consider the expected value
\[
\mathrm{E}[W_i\cos\theta_i]=\mathrm{E}[(X_i\cos\theta_i+Y_i\sin\theta_i)\cos\theta_i].
\]
Because $\theta_i$ is independent of $X_i$ and $Y_i$,
\begin{eqnarray*}
&\mathrm{E}[W_i\cos\theta_i] & =  \mathrm{E}[\mu_{xi}\cos^2\theta_i+\mu_{yi}\sin\theta_i\cos\theta_i] \\
 &= & \int_0^{2\pi} \left(\mu_{xi}\cos^2\theta_i+\mu_{yi}\sin\theta_i\cos\theta_i\right)P(\theta)\mathrm{d}\theta \\
 &= & \mu_{xi}/2.
\end{eqnarray*}
By the Law of Large Numbers, we can estimate $\mu_{xi}$ with
\[
\widehat{\mu_{xi}}=2\overline{W_i\cos\theta_i}=\frac{1}{n}\sum_{k=1}^nW_{i}^{(k)}\cos\theta_i^{(k)},
\]
where we use the hat to denote the estimate of a parameter and the overline to denote the sample mean.  Similarly, we can estimate $\mu_{yi}$ with $\widehat{\mu_{yi}} = 2\overline{W_i\sin\theta_i}$. Applying this treatment to both modes gives us $\widehat{\mu}$.

To estimate $\Sigma$, consider the expected value
\[
\mathrm{E}[W_i^2\cos^2\theta_i]=\mathrm{E}[(X_i\cos\theta_i+Y_i\sin\theta_i)^2\cos^2\theta_i].
\]
Using the independence of $\theta_i$ from $X_i$ and $P_i$ and the uniformity of $P(\theta_i)$, we obtain
\begin{eqnarray*}
\mathrm{E}[W_i^2\cos^2\theta_i] & =&\mathrm{E}\left[\frac{1}{8}(3X_i^2+Y_i^2)\right] \\
 {} & =& \frac{1}{8}(3\mu_{xi}^2+\Sigma_{xi,xi}+\mu_{yi}^2+\Sigma_{yi,yi}).
\end{eqnarray*}
One can similarly show that
\[
\mathrm{E}[W_i^2\sin^2\theta_i] = \frac{1}{8}(\mu_{xi}^2+\Sigma{xi,xi}+3\mu_{yi}^2+3\Sigma_{yi,yi}),
\]
and
\[
\mathrm{E}[W_i^2\cos\theta_i\sin\theta_i] = \frac{1}{4}(\mu_{xi}\mu_{yi}+\Sigma_{xi,yi}).
\]
Solving these three equations for the elements of $\Sigma$ and applying the Law of Large Numbers gives us the estimates
\begin{eqnarray*}
\widehat{\Sigma_{xi,xi}} & =&3\overline{W_i^2\cos^2\theta_i}-\overline{W_i^2\sin^2\theta_i}-\widehat{\mu_{xi}}^2 \\
\widehat{\Sigma_{yi,yi}} & =&3\overline{W_i^2\sin^2\theta_i}-\overline{W_i^2\cos^2\theta_i}-\widehat{\mu_{yi}}^2 \\
\widehat{\Sigma_{xi,yi}} & =&4\overline{W_i^2\cos\theta_i\sin\theta_i}-\widehat{\mu_{xi}}\widehat{\mu_{yi}}.
\end{eqnarray*}
Estimates for the cross-mode elements of $\Sigma$ are given by
\begin{eqnarray*}
\widehat{\Sigma_{x1,x2}} & =&4\overline{W_1\cos\theta_1W_2\cos\theta_2}-\widehat{\mu_{x1}}\widehat{\mu_{x2}} \\
\widehat{\Sigma_{x1,y2}} & =&4\overline{W_1\cos\theta_1W_2\sin\theta_2}-\widehat{\mu_{x1}}\widehat{\mu_{y2}} \\
\widehat{\Sigma_{y1,x2}} & =&4\overline{W_1\sin\theta_1W_2\cos\theta_2}-\widehat{\mu_{y1}}\widehat{\mu_{x2}} \\
\widehat{\Sigma_{y1,y2}} & =&4\overline{W_1\sin\theta_1W_2\sin\theta_2}-\widehat{\mu_{y1}}\widehat{\mu_{y2}}.
\end{eqnarray*}

Thus, we estimate the expected values $\mu$ and covariance matrix $\Sigma$ by computing sample means of simple functions of the quadrature and phase measurements.  Because the computation is so simple, our method is well suited for large data sets.  The maximum likelihood method developed by \v{R}eh\'{a}\v{c}ek and co-authors might give lower statistical uncertainty at the cost of greater computation time \cite{Rehacek2009}. Our method does not impose a constraint on $\Sigma$ to enforce the Heisenberg Uncertainty Principle \cite{Simon1994}. Although it is possible for the estimate to be unphysical, all states found in our analysis are valid quantum states.

\begin{acknowledgments}
The authors acknowledge support from the DARPA/MTO QuEST program. This work is supported by the DARPA/MTO QuEST program and National Science Foundation under Grant Number 1125844.
\end{acknowledgments}

\bibliography{mybib}

\end{document}